\documentclass[12pt,preprint]{aastex}

\shorttitle{On the Reality of the Accelerating Universe}
\shortauthors{A. M\'esz\'aros}

\begin{document}
 
\title{On the Reality of the Accelerating Universe}

\author{Attila M\'esz\'aros}

\affil{Astronomical Institute of the Charles University,
              V Hole\v{s}ovi\v{c}k\'ach 2, 180 00 Prague 8,
	      Czech Republic}

\begin{abstract}
   Two groups recently deduced the positive value for the cosmological 
constant, concluding at a high ($\geq 99\%$) confidence level that the 
Universe should be accelerating. This conclusion followed from the 
statistical analysis of dozens of high-redshift supernovae. In this paper 
this conclusion is discussed. From the conservative frequentist's point of 
view the validity of null hypothesis of the zero
cosmological constant is tested by the classical statistical $\chi^2$ test
for the 60 supernovae listed in Perlmutter et al. 1999 (ApJ, 517, 565).
This sample contains 42 objects discovered in the frame of Supernova Cosmology
Project and 18 low-redshift object detected earlier. Excluding the event 
SN1997O, which is doubtlessly an outlier, one obtains the result: The
probability for seeing a worse $\chi^2$ - if the null hypothesis is
true - is in the $5\%$ to $8\%$ range, a value that does not
indicate significant evidence againts the null.
If one excludes further five possible outliers, proposed to be done by 
Perlmutter et al. 1999, then the sample of 54 supernovae 
is in an excellent accordance with the null hypothesis. It also seems that 
supernovae from the High-$z$ Supernova Search Team does not change
the acceptance of null hypothesis. This means that
the rejection of the Einstein equations with zero cosmological constant -
based on the supernova data alone - is still premature.
\end{abstract}


\section{Introduction}

In recent years two independent groups (\citet{pe99}, \citet{ri00a}
and references therein) concluded that the cosmological constant
is positive with $\Omega_{\Lambda} \simeq 0.7$ and 
$\Omega_M \simeq 0.3$ (for a detailed review 
and further references see, e.g., \citet{ri00b}; for the latest
developments see \citet{ri01}).
As usual, $\Omega_M$ denotes the ratio of the density of 
the non-relativistic matter in Universe
to the critical density; $\Omega_{\Lambda} = \lambda c^2/(3H_o^2)$,
where $\lambda$ is the cosmological constant, $c$ is the velocity of light,
and $H_o$ is the Hubble constant.  
This conclusion was based purely on the data of the observations done in the
Supernova Cosmology Project (\citet{pe99} and references therein)
and of the observations done by the High-$z$ Supernova Search Team
(\citet{sch98}, \citet{ri00a} and re\-ferences therein). The Universe
should also be accelerating, because $\Omega_{\Lambda} >
\Omega_M/2$ (see \citet{ri00b} for more details). 

Both teams recognize that the supernovae at redshift $z \simeq (0.3 - 1.0)$ 
give in average 
a $\simeq 0.28$ mag bigger distance moduli than expected if
$\Omega_M \simeq 0.3$ and $\Omega_{\Lambda} = 0$ \citep{ri00b}.
This excess of distance modulus
is so small and there are so many sources of uncertainties that
extreme care is needed in drawing conclusions. This fact is, of course,
clearly proclaimed by both teams. Therefore, 
further careful analysis concerning the methods, statistics, errors, 
alternative explanations, etc... are highly required. Any new result - 
even of minimal technical importance - is highly desirable 
and should immediately be announced (R. Kirshner, private communication).
For example, \citet{dre00} and \citet{go01} gave smaller evidence for the
non-zero cosmological constant.

This article is - in essence - also such a required contribution. 
It discusses one concrete question of the
topic; namely, the probability of the rejection 
of the zero cosmological constant hypothesis. The discussion is done
from the pure statistical point of view.

\section{General considerations}

In \citet{pe97} and \citet{pe99} (in what follows P99)
their analysis of the data gives the
conclusion that  
$\Omega_{\Lambda} > 0$ holds with a $99\%$ confidence. 
In \citet{ri00a} a higher than $99\%$ confidence is 
deduced. \citet{go01} 
deduced - from an earlier sample - that the confidence
for $\Omega_{\Lambda} > 0$ is only $89.5\%$ or smaller.
All these statistical analyses followed the so called "Bayesian approach".
The key idea of this approach is based on the procedure, in which - even before
existing any measured data concerning a hypothesis - some preliminary
degree of plausability ("Bayesian prior") is assigned to the hypothesis
(for more details about the Bayesian approach in Astronomy
see, cf., \citet{dre00} and references therein; about the different aspects
of methods from the statistical point of view see \citet{be02}).
In the case of the supernovae the different confidence levels came
from the different prior of the hypothesis $\Omega_{\Lambda} = 0$.

The author means that it is highly useful to provide an analysis
of data from the frequentist's point of view, too. This
approach proceeds classically and most conservatively. This means
that - at the begining - it is simply assumed that the Friedmannian model
(either with $\Omega_M > 1$ or with with $\Omega_M = 1$ or
with $0< \Omega_M < 1$) {\it with zero cosmological constant}
is the correct model.
Then it is asked that the observational data are in accordance with
this model or not (for more details concerning this statistical
approach see any standard text-book of Statistics
\citep{tw53,ks76}; from newer publications see, e.g., \citet{fc98}
and references therein). 

The requirement of this analysis may be supported as follows.

It is a standard knowledge that the Friedmannian model with
$\Omega_{\Lambda} = 0$ is based
on two different assumptions: A. Gravitation is described by the
Einstein equations with zero cosmological constant. B. The Universe
has a symmetry defined by six linearly independent Killing vectors, and
the character of this symmetry allows to speak about a maximally symmetric
three-dimensional sub-manifold 
(this assumption is called as "cosmological principle";
for more details see, cf. \citet{we72}; Chapt. 14.1).

The verification of these assumptions should be done by observations, 
of course. In the verification of
the assumption B. it is quite usual to proceed in the frame of the
most conservative point of view. In addition, theoretically,
even if the cosmological principle were rejected, it would not be clear, 
which non-Friedmannian model should then 
be used (for the survey of non-Friedmannian models
see, cf., \citet{kra97}). Simply, if the cosmological principle
is not rejected yet unambiguously at a high significance, then the
best is "to keep the cosmological principle as far as possible". 

The author means that one should similarly proceed concerning
the assumption A., too. From the observational
point of view \citet{dre00} and \citet{go01} suggest that one
should remain careful in the final conclusions. In addition,
from the theoretical point of view, even if the observations
were rejecting assumption A., one would be able to introduce {\it several
different generalizations} of Einstein equations. For example, \citet{go01}
discusses both the usual generalization with cosmological constant, but
also the possibility with the 
time-variable "constant". This second possibility
is identical to the introduction of a long-range scalar field coupled with
the gravitation. In fact,
there are known many similar theoretical attempts for other fields
coupled with the gravitation (see \citet{go01} and references therein).
Add here that also the author probed to introduce such long-range force
defined by a pair of standard spin-2 fields \citep{me87}. This probe was
proclaimed to be hopeless, because of the unsolvable complications
in the theory \citep{me91}. 
In any case, the introcution of non-zero cosmological constant is not the only
possible generalization of Einstein equations. 
Simply, also here the best choice is
"to keep the assumption A. as far as possible".

The observations quite unambiguously suggest that $\Omega_M
\geq 1$ is excluded and, in addition,
 it may be assumed $0.1 \leq \Omega_M \leq 0.5$
\citep{bf98}. Hence, the null hypothesis
will be the assumption of the correctness of the Friedmannian model with
$0.1 \leq \Omega_M \leq 0.5$ and $\Omega_{\Lambda} = 0$. The sample obtained
from observations will be given by 
the 60 supernovae collected and discussed at P99.
 Hence, the precise purpose of this article is to test:
Does this sample alone reject the null hypothesis? This is studied in
Section 3. The remaining supernovae from the second project, 
and also some other questions, will shortly be discussed in Section 4.
In Section 5 the results of paper will be summarized.

\section{The $\chi^2$ test}

Be given the data of 60 supernovae collected at
Table 1 and Table 2 of P99. Then
one has to fit the $[m_B^{eff}, \log z]$ data-pairs with
the theoretical curves, in which $\Omega_{\Lambda} = 0$ holds
{\it identically}. 
This means that there are only two independent parameters in these theoretical
curves ($H_o$ and $\Omega_M$). The procedure is a standard one, 
and is described, e.g., by \citet{pre92}, Chapt.15.1. One has to
do three things: To determine the two best-fit parameters; to determine
their allowed ranges; to determine the goodness-of-fit due to the standard
$\chi^2$ test. 
Eq. 15.1.5 of \citet{pre92} takes the form
\begin{equation}
\chi^2 = \sum_{i=1}^{N} \Bigl(\frac {m_{Bi}^{eff} - 
m_B^{eff}(z_i, H_o, \Omega_M, \Omega_{\Lambda} =0)}{\sigma_i}\Bigr)^2,
\end{equation}
where $N$ is the number of supernovae in the sample, and $\sigma_i$ is the 
uncertainty of the effective magnitude of $i$-th supernova having measured 
the redshift $z_i$ and corrected $B$ band magnitude 
$m_{Bi}^{eff}$ ($i=1,2,...,N$). The corrected $B$ band
magnitude is given by
\begin{equation}
m_B^{eff} = 25 + M_B + 5 \log (c/H_o) + 
5 \log Q(z, \Omega_M, \Omega_{\Lambda} = 0),
\end{equation}
where $M_B$ is the absolute magnitude, $c/H_o$ is in Mpc,
and one has \citep{car92}
\begin{equation}
Q(z) = (2/\Omega_M^2)
(2 + \Omega_M (z-1) - (2 - \Omega_M)\sqrt{1+\Omega_M z} ).
\end{equation}
This standard relation of Cosmology
is obtainable also directly \citep{mm96} without the integration of
general equation presented by \citet{car92}. 
The null hypothesis should then 
be rejected in the case when either the best-fit parameters are fully
unphysical or the goodness-of-fit excludes the fit itself. In our case
we will proceed in such a way that only the observationally allowed
ranges of parameters will be considered - hence, if one obtains a good fit
from the goodness-of-fit, then the fit is immediately acceptable. 

In our case $N=60$, and one may take in accordance with \citet{pe97}
$\tilde{M} = M - 5 \log H_o + 25 = -3.32$.
A $\simeq 12\%$ observational uncertainty in the value of 
$H_o$ \citep{go01,fr01} 
gives maximally a $\simeq 5 \log 1.12 = 0.28$ mag change in the value
of $\tilde{M}$. This means that in the range
$3.60 > - \tilde{M} > 3.04$ one should search for the best fits.

Using the measured
redshifts, corrected effective magnitudes and their uncertainties one
obtains the best fit for $\Omega_M = 0.1$, and for $\tilde{M} = 3.30$;
namely $\chi^2 = 107.1$. This value is the best fit for
58 degrees of freedom. 

Varying the free parameters in the allowed ranges one obtains the following.
If $\tilde{M} = -3.32$ and $\Omega_M = 0.1$, one has
$\chi^2 = 108.0$. In fact, in all fits of this Section the best fits for
$\tilde{M}$ were practically always given by $\tilde{M} = -3.32$. 
The values of $\chi^2$ obtained for $\tilde{M} = -3.32$ and for the best
fits values of $\tilde{M}$ gave practically the same significance
levels - the differences were smaller than $1\%$, which is unimportant
for the purpose of this paper. Therefore, in
what follows, the value of $\tilde{M} = - 3.32$ may always be taken as the
best fit value. In the case of parameter $\Omega_M$
the worst fit is obtained for $\Omega_M = 0.5$, namely 
$\chi^2 = 129.6$. Between $\Omega_M = 0.5$ and $\Omega_M = 0.1$
the fitting is monotonously strengthening, if one goes toward the smaller
values. Contrary to $\tilde{M}$, in the case of parameter 
$\Omega_M$, the best fit value is on the boundary 
of the allowed range of parameter.

The goodness-of-fit is given by the chi-square probability function
$P(\nu/2, \chi^2/2)$
(cf. \citet{pre92}, Chapt.6.2) for $\nu = 58$ degrees of freedom.
Add here that fast approximate probability of the goodness-of-fit is
obtainable also without the calculation of this function directly from
the table of $\chi^2$ distribution (see \citet{tw53}, Table A5).
One may use the fact that roughly
for $\nu > 20$ degrees of freedom the reduced 
$\chi^2/\nu$ distribution is practically not changing.
For $\nu = 58$ and $\chi^2 = 108$ the significance level is
between $1\%$ and $0.1\%$. For $\Omega_M = 0.5$ the fit even is worse, and
the significance level is around $0.1\%$. 

For the sake of statistical precisity two notes must be added here.

The first one concerns the degrees of freedom. In fact, $m_B^{eff}$ itself is
a corrected value in P99, and contains two further parameters (see P99
for more details). Hence, the degree of freedom, as it seems, for 60 objects
should be $\nu = 56$. In 
fit A of Table 3 in P99 this values is used.
Further complications can come
from the fact that the best fit value $\Omega_M = 0.1$ is a
boundary value of allowed range. This may cause
some problems (for a discussion of this question
see, for example, \citet{pro02}). In our case 
this may mean that - in essence - 
$\Omega_M$ should not be considered as a free
parameter, but the value $\Omega_M = 0.1$ should be fixed immediately.
In this case the degree of freedom should be increased by one.
All this means that there is an ambiguity in the
concrete value of the degree of freedom.
Furtonately, in our case, this problem is not essential:
The significances are the same - the difference is smaller than $1\%$, once
the degree of freedom is changed by one. Therefore, in what follows, we
may further take $\nu = N-2$ for $N$ objects.

The second notes concerns the errors. Strictly speaking, one should include
into $\sigma_i$ also the errors of $\log z$ (see \citet{pre92}, Chapt. 15.3).
But the errors in $\log z$ - compared with the errors of magnitudes -
are small (except for some low redshift objects). In any case, these
additional errors should decrease the significances of rejection, because
they should decrease the value of $\chi^2$.
But this effect should also be unimportant here. 

The approximate significance from the reduced $\chi^2/\nu$,  
the effect of errors in redshifts, the boundary value of $\Omega_M$ together
with the ambiguity in the degree of freedom, and
the choice $\tilde{M} = -3.32$
may cause a maximally $(1-3)\%$ uncertainty in the obtained significance.
This inpreciseness is inessential for our purpose.

For our purpose it is essential that the sample with $N = 60$ supernovae gives
a fully wrong fit. {\it The null hypothesis for the whole sample 
should be rejected; the significance level is in the
range $(0.1 - 3.0) \%$, being enough to reject the null.}

Nevertheless, doing a final conclusion, a care is still needed. This follows
from the following fact. An inspection of terms in 
$\chi^2 = 108$ for $\Omega_M =0.1$ shows that 
in this sum a big amount, namely $26.7$, is given by one supernova, 
namely by SN1997O having $z=0.374$. Hence, if this one single
object were not considered in the sample, then the sample with $N=59$
object would give only $\chi^2 = 81.3$ for $\nu = 57$ 
degrees of freedom. This would 
already be an acceptable fit, because the rejection of null hypothesis
would occuring at a $(6-7)\%$ significance level. 
Taking into account also the possible $(1-3)\%$ uncertainty,
one may conclude that the the usually
requested $<5\%$ significance level - 
allowing the rejection - should not be reached.

Hence, we arrive at the surprising result: 
{\it The null hypothesis is rejected; but by one single object!}

There are three different arguments suggesting that
SN1997O should actually be removed from the sample.

The first argument comes from general statistical considerations
of outliers. It is never strange in Statistics to remove
an object from the sample, if this is an "outlier". Generally, outliers are
observations which are inconsistent with the remainder of the data set
(a detailed discussion of outliers is given, e.g., by \citet{jo86}; 
Chapt.10.1). Looking into Figures 1 and  2 of P99, one immediately
sees that just SN1997O is a good candidate for an outlier, because
it is far above the magnitudes expected from the
trend given by other objects. The object is "too faint".

The second argument follows from the text of P99. This article
also discusses the question of outliers from astrophysical point of view
(different light curves, reddening in the host galaxy, etc.). Four
supernovae, namely SN 1992bo, 1992bp, SN1994H, SN1997O, 
are proclaimed as "most
significant outliers". There are further two ones (SN1996cg,
SN1996cn), which are also proposed
not to be taken into the sample from different reasons. Hence,
also P99 takes SN1997O as an outlier, too.
In addition, further three or five objects are proposed to be removed,
too.

The third argument follows the following consideration. Assume that no
outliers are in the sample. Then the null hypothesis is rejected, and
the generalization of Einstein equations is needed. 
There are several possibilities for this generalization. One of this is
the non-zero cosmological constant. Then one should fit the whole sample
with $N=60$ 
with the theoretical curves allowing $\Omega_{\Lambda} \neq 0$. This was
already done by P99 (Table 3, fit A); the value $\chi^2 = 98$ was obtained.
But this is {\it again a wrong fit}, 
because for 58 degrees of freedom one obtains a
{\it rejection} at the significant level $1\%$ \citep{tw53}. All this means
that in this case {\it both $\Omega_{\Lambda} = 0$
and $\Omega_{\Lambda} \neq 0$ should be rejected}. 
Simply, also the generalization
with non-zero cosmological constant is {\it not} acceptable. 
It is even questionable that any theoretical curve - in the frame of
cosmological principle - can fit this sample (see \citet{we72}, Chapt. 14
for the general discussion of theoretical curves). Hence,
the object SN1997O {\it alone} should reject the Einstein equations both with
zero and non-zero cosmological constant; in addition, probably also
the cosmological principle itself.

The author means - in accordance with P99 - that this object
is a clear outlier and should be removed from the sample.
All this means that the best is to consider three different samples. The first
one is the sample with $N=59$ objects removing only SN1997O. 
The second 
sample is the sample B of P99; the third one is the "primary sample" of
P99 having $N=54$ objects (sample C). 
P99 proposes to use this third sample as the best primary choice.

The first sample with $N=59$ 
gives an acceptable fit; the rejection of null hypothesis
should be at the significance level $(5-8)\%$.
The second sample with $N=56$ 
gives $\chi^2 = 68.3$ for $\Omega_M = 0.1$, which is
again an acceptable fit for $\nu =54$ degrees of freedom. The null 
hypothesis is rejected at $11\%$ significance level.
The primary sample with $N=54$ 
gives $\chi^2 = 63.7$ for $\Omega_M = 0.1$. This
value for $\nu = 52$ degrees od freedom gives an excellent fit - the
null hypothesis is rejected at the $28\%$ significance level.
In any case, $<5\%$ level is never reached. 

\section{Discussion}

In \citet{ri98} (see also \citet{ri99})
there are discussed 10 further high-redshift
supernovae with $0.16 \leq z \leq 0.62$. Of course, the best solution
 would be to fit these objects together with the 60 objects of P99. 
Nevertheless, the errors in \citet{ri98} are listed in other way than in P99. 
In addition -
even having a list of $\sigma_i$ for any object obtained by the same
manner - further complications can arise from the existence of outliers.
(Clearly, the same criterion for an outlier should be required
in such a "matched" sample. It is not clear, how to define this criterion.)
Simply, the matching of the all possible observed supernovae into one single
statistical sample leads to several technical problems, and the author
- not being in the teams of two projects - is not able to solve this
technical question. Therefore, here, the supernovae from the second 
team will be fitted separately. 

Using Eq.4 of \citet{ri98}, in which $\Omega_{\Lambda} = 0$ and
$\sigma_{v} = 0$, one may provide the fitting for 10 objects listed in
Table 5 and Table 6 of \citet{ri98}. Taking the values of $\mu_o$ and
$\sigma_{\mu_o}$ from the last column of Table 5, and taking the possible
values of free parameter $H_o$ (in units km s$^{-1}$ Mpc$^{-1}$) between 
64 and 80 \citep{fr01}, one obtains the best fit for $H_o = 79$ 
km s$^{-1}$ Mpc$^{-1}$, and $\Omega_M = 0.1$; namely $\chi^2 = 9.03$.
Taking the values of $\mu_o$ and
$\sigma_{\mu_o}$ from the last column of Table 6, 
one obtains the best fit for $H_o = 78$ 
km s$^{-1}$ Mpc$^{-1}$, and $\Omega_M = 0.1$; namely $\chi^2 = 7.7$.
Both cases are excellent fits, because the significance level is
around $40\%$ and $50\%$, respectively,
for 8 degrees of freedom \citep{tw53}. Note here that
the change caused by $\Omega_M$ is weak. For example, in the first case
for $\Omega_M = 0.5$ one still has $\chi^2 = 9.6$. The dependence on
the change of $H_o$ is more essential, but in any case for $H_o = 77-80$ 
km s$^{-1}$ Mpc$^{-1}$ acceptable fits are obtained. The choice
$\sigma_{v} = 0$ is not a problem, because eventual non-zero values
further decrease the value of $\chi^2$ and thus further strengthen the goodness
of fits. The 10 supernovae from \citet{ri98} 
{\it alone} do not need non-zero cosmological constant.
 
For the sake of completeness the object SN1997ff with $1.5 < z
< 1.8$ should also be discussed \citep{ri01}. For this object the uncertainty
at $\Delta(m-M)$ is so large ($\simeq 1$ mag, as this
is clear from Figures 10 and
11 of \citet{ri01}) that here the contribution for $\chi^2$ should
surely be smaller than 1. This object alone 
should even strengthen the acceptance of null hypothesis.

Discussing the results of article
it must still be precised the following. Strictly speaking, this article
does not claim that the introduction of non-zero cosmological constant
cannot be done. It is only said that - purely from the most conservative
statistical point of view and purely from the supernovae
observational data {\it alone} - the assumption of zero cosmological
is {\it not} rejected yet at a high enough significance level. 
The reality of the non-zero cosmological
constant is not excluded; it remains an open problem yet.
In any case, the different statistical methods - either
from the Bayesian \citep{dre00,go01} or from frequentist's point of view
(this article) - still suggest that - from the statistical point of
view  - the "definite", "final"
or "unambiguous" introduction of non-zero cosmological term - based on
the supernova data alone - 
is still premature. In fact, this is the key result of this article. 

\section{Conclusions}

The results of paper may be summarized as follows.

   \begin{enumerate}
      \item The observational data of 60 supernovae - listed in P99 - were 
reanalyzed from the conservative statistical point of view. 
The null hypothesis of zero cosmological constant is not rejected by these
data alone.
The probability for seeing a worse $\chi^2$ - if the null hypothesis is
true - is in the $5\%$ to $28\%$ range, a value that does not
indicate significant evidence againts the null.
If only one clear outlier is omitted, then this 
probability is $(5-8)\%$; if further outliers - proposed by
P99 - are omitted, then this probability is $(10-28)\%$.
The value $<5\%$ is not reached.
      \item The High-z Supernova Search Team data alone
suggest that this conclusion further holds.
      \item All this means that the introduction of non-zero 
cosmological constant - based on the supernovae data alone -
is still premature. The reality of the non-zero
cosmological constant remains an open question.
   \end{enumerate}

\acknowledgments
The author thanks the valuable discussions with
L.G. Bal\'azs, A. Graham, R. Kirshner, J. Lub and M. Wolf. 
The useful remarks of an anonymous referee are kindly acknowledged.
This research was supported by Czech Research Grant
J13/98: 113200004.

\end{document}